\documentclass[12pt]{iopart}
\usepackage{iopams}  
\usepackage{graphicx}
\usepackage{appendix}
\usepackage{color} 
\usepackage{cite}
\usepackage[position=top,singlelinecheck=off,justification=raggedright,caption=false]{subfig}
\usepackage{epstopdf} 
\DeclareGraphicsExtensions{.pdf,.eps,.png,.jpg,.mps}
\usepackage[colorlinks=true,citecolor=blue,urlcolor=blue,linkcolor=blue]{hyperref}
\usepackage{bm}
\usepackage{dcolumn}

\newcommand{\ket}[1]{\left|#1\right\rangle}
\newcommand{\braket}[2]{\left\langle#1 |  #2\right\rangle}

\def\BEq{\begin{equation}}
\def\EEq{\end{equation}}
\def\BEqA{\begin{eqnarray}}
\def\EEqA{\end{eqnarray}}
\def\BW{\begin{widetext}}
\def\EW{\end{widetext}}

\begin{document}

\title[]{Quantum simulation of macro and micro quantum phase transition from paramagnetism to frustrated magnetism with a superconducting circuit}

\author{Joydip Ghosh}
\ead{ghoshj@ucalgary.ca}
\address{Institute for Quantum Science and Technology, University of Calgary, Calgary, Alberta T2N 1N4, Canada}
\author{Barry C. Sanders}
\ead{sandersb@ucalgary.ca}
\address{Institute for Quantum Science and Technology, University of Calgary, Calgary, Alberta T2N 1N4, Canada}
\address{Program in Quantum Information Science, Canadian Institute for Advanced Research, Toronto, Ontario M5G 1Z8, Canada}
\address{Hefei National Laboratory for Physical Sciences at the Microscale and Department of Modern Physics, University of Science and Technology of China, Anhui 230026, China}
\address{Shanghai Branch, CAS Center for Excellence and Synergetic Innovation Center in Quantum Information and Quantum Physics, University of Science and Technology of China, Shanghai 201315, China}

\date{\today}

\begin{abstract}
We devise a scalable scheme for simulating a quantum phase transition
from paramagnetism to frustrated magnetism in a superconducting flux-qubit network,
and we show how to characterize this system experimentally
both macroscopically and microscopically.
The proposed macroscopic characterization of the quantum phase transition
is based on the transition of the probability distribution
for the spin-network net magnetic moment
with this transition quantified by the difference between the Kullback-Leibler divergences of the
distributions corresponding to the paramagnetic and frustrated magnetic phases 
with respect to the probability distribution at a given time during the transition.
Microscopic characterization of the quantum phase transition is performed using the standard
local-entanglement-witness approach.
Simultaneous macro and micro characterizations of quantum phase transitions
would serve to verify a quantum phase transition in two ways 
especially in the quantum realm for the classically intractable case of frustrated quantum magnetism.
\end{abstract}

\vspace{2pc}

\maketitle

\section{Introduction}
\label{sec:introduction}

Experimental quantum simulation~\cite{GAN14} opens vistas for exploring foundational quantum principles
such as studies of Quantum Phase Transitions (QPT)~\cite{sach2011,Greiner2002,Soltan-Panahi2012,Baumann2010,MBK+15}.
Although the \emph{phase} of matter can only be defined macroscopically,
the QPTs are currently explored through microscopic characterizations
that depend on local observables such as individual-particle spin states or 
few-body states or quantum entanglement witnesses~\cite{kim2010quantum}.
Macroscopic probing, which measures some global property of a system 
without having any access to individual particles, 
is not currently employed in quantum-phase-transition experiments.
Yet conducting joint macro and micro characterizations of quantum phases and transitions between them
is important for self-consistent verification of QPTs in the macro and micro regimes, 
assuming that one may not necessarily imply another~\cite{Anderson04081972, Kle67}.

Here we propose a scalable scheme comprising nearest- 
and next-nearest-neighbor-coupled spin-half network of 
radio-frequency superconducting-quantum-interference-device
(rf-SQUID) flux qubits~\cite{HJB+10,PhysRevB.82.024511}, 
where the two levels comprise one clockwise and one counter-clockwise super-currents.
The clockwise super-current state~$\ket{\circlearrowright}$
is equivalent to a spin-down flux state~$\ket{\downarrow}$,
and the counter-clockwise super-current state~$\ket{\circlearrowleft}$
corresponds to a spin-up flux state~$\ket{\uparrow}$.
This flux-qubit spin network will realize,
and admit macro and micro characterization of,
a controllable QPT from paramagnetism to frustrated magnetism~\cite{diep2004frustrated}.

We focus on quantum simulation of the paramagnetic to frustrated-magnetic
QPT~\cite{MBK+15} because of the need to simulate
the fascinating behavior manifested in frustrated spin systems~\cite{diep2004frustrated},
such as spin glasses, spin ices and quantum spin liquids
due to competing spin-spin interactions.
These competitions give rise to an
exponentially large subspace of degenerate ground states for such frustrated networks~\cite{doi:10.1021/ja01315a102},
which strongly motivates quantum simulation of these systems~\cite{GAN14}.
We seek to identify the macro signature
of the onset of quantum frustrated magnetism 
from paramagnetism through the statistics of the order parameter rather than solely by its mean.

Quantum simulations of frustrated Ising networks~\cite{Ong02072004}
have been performed with ion traps~\cite{kim2010quantum},
quantum dots~\cite{PhysRevLett.110.046803},
and nuclear magnetic resonance~\cite{PhysRevA.88.022312}
and has been proposed for Rydberg atoms~\cite{GDN+15}.
Larger frustrated systems have been realized for optical systems~\cite{ma2011quantum}
and trapped ions~\cite{Islam03052013}.
All these schemes employ microscopic probes of frustration
and are not yet amenable to macroscopic probing,
which is why we propose a new setup 
corresponding to a scalable flux-qubit superconducting architecture
that admits probing of both macro and micro signatures of frustration.

The rest of the paper is organized as follows: Section~\ref{sec:model} describes the physical model for the architecture considered in this work. In Section~\ref{sec:simulation} we elaborate the control procedure and discuss how micro and macro signatures can be extracted for our system. Section~\ref{sec:results} shows the results of our simulation and we conclude in Section~\ref{sec:conclusions}.

\section{Physical model}
\label{sec:model}

\begin{figure}
\centering
\subfloat[]{
\centering
\includegraphics[width=0.6\linewidth]{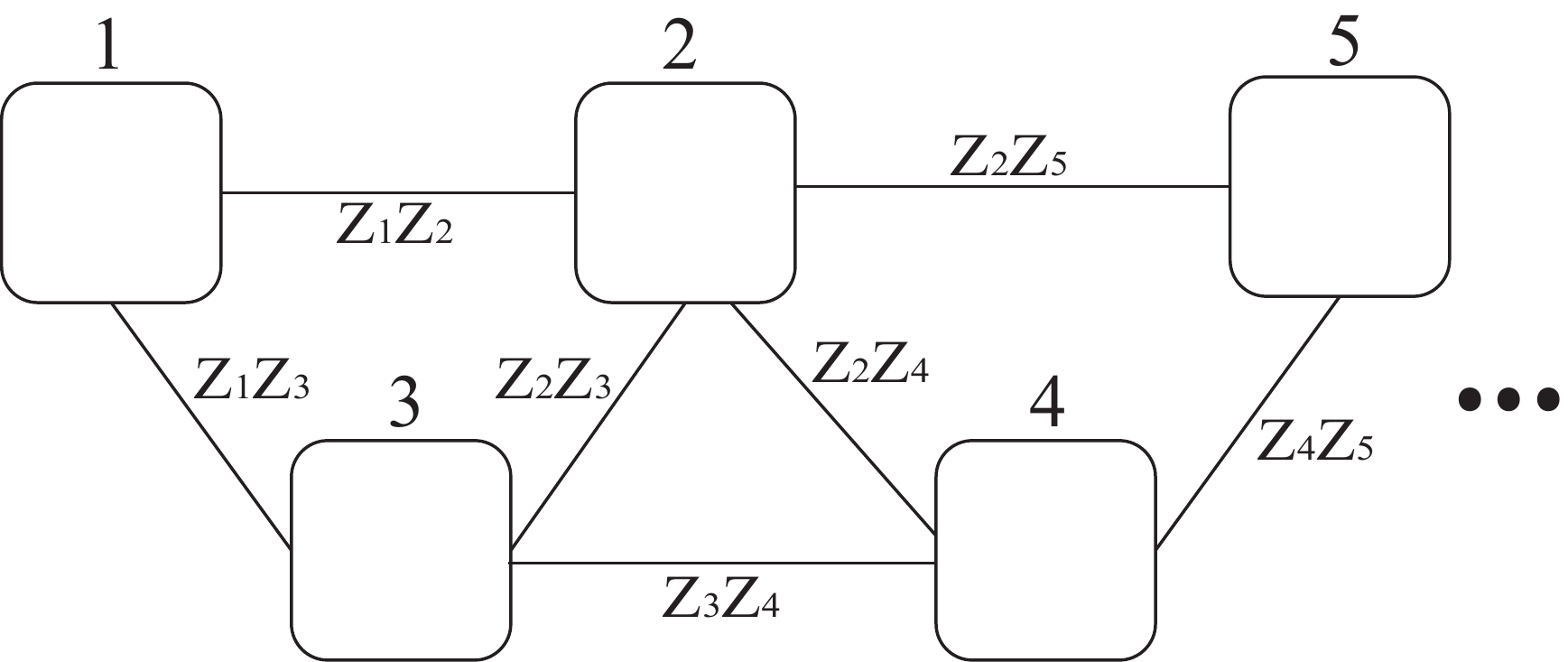}
\label{fig:frustrationChain}
} \\
\subfloat[]{
\centering
\includegraphics[width=0.35\linewidth]{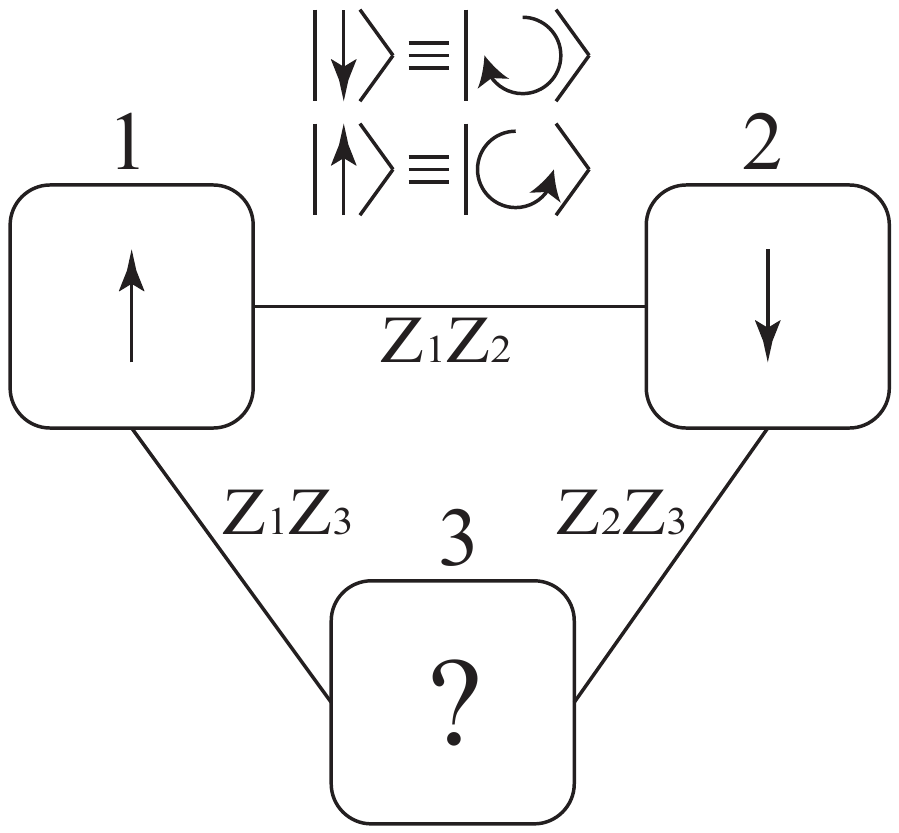}
\label{fig:frustrationScheme}
}
\caption{%
	Schematic for (a)~$n$ nearest-neighbor- and next-nearest-neighbor-coupled qubits and 
	(b)~three coupled qubits in a triangular architecture, which is the smallest lattice required 
	for frustrated magnetism. Rounded squares denote qubits, and solid lines denote couplings.}
\end{figure}

Our goal is to analyze the model for nearest- and next-nearest-neighbor-coupled rf-SQUID superconducting flux qubits (as shown in Fig.~\ref{fig:frustrationChain}) and devise a scheme to extract the macroscopic as well as microscopic signatures of QPT from paramagnetism to frustrated magnetism. However, we first consider three rf-SQUID superconducting flux qubits coupled to each other via tunable inductive couplings (shown in Fig.~\ref{fig:frustrationScheme}), as this triangular architecture is the smallest lattice that can achieve the regime of frustrated magnetism.

\subsection{CCJJ flux qubit}
\begin{figure}
\centering
\includegraphics[width=0.5\linewidth]{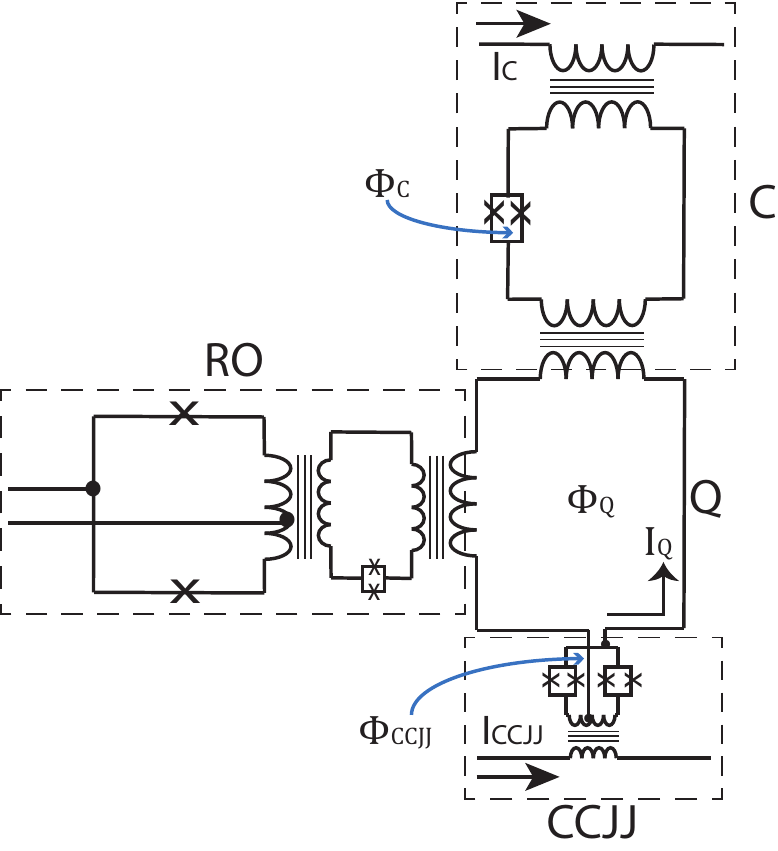}
\caption{%
	A circuit diagram of an rf-SQUID CCJJ superconducting flux qubit, 
	where the quantized circulating current $I_{\rm Q}$ across the qubit loop 
	(denoted by `Q') defines the $\ket{\circlearrowright}$ and $\ket{\circlearrowleft}$ states of the qubit.
	The CCJJ loop comprising four Josephson junctions is coupled to a bias current~$I_{{\rm CCJJ}}$
	for controlling the height of the flux-qubit potential-energy barrier.
	The time-dependent bias current $I_{{\rm CCJJ}}(t)$
	induces a time-dependent magnetic flux $\Phi_{{\rm CCJJ}}(t)$ across the CCJJ loop.
	This flux controls the $\ket{\circlearrowright}\leftrightarrow\ket{\circlearrowleft}$ tunneling probability.
	The compensator~C,
	which is biased by current $I_{{\rm C}}$ via an in situ tunable inductive coupler to
	the flux qubit,
	is controlled by the static flux bias $\Phi_{{\rm C}}$. It compensates for the changing current $I_{{\rm Q}}$, such that the qubit frequency~$\epsilon$ can be maintained to a specific value during the experiment.
	The readout circuitry (denoted `RO') can detect any variation of the circulating current $I_{{\rm Q}}$ and is used to perform the readout operation on the flux qubit.
}
\label{fig:CCJJ}
\end{figure}

The Compound-Compound Josephson-Junction (CCJJ)~$\ket{\circlearrowright}$
shown in Fig.~\ref{fig:CCJJ}
is proposed here as our favored flux qubit 
(with basis states~$\ket{\circlearrowright}$ and~$\ket{\circlearrowleft}$)
due to its exquisite tunability~\cite{PhysRevB.82.024511,HJB+10}
and its demonstrable scalability in the sense that hundreds of such flux
qubits can be coupled in a spin network~\cite{Dickson2013,PhysRevX.4.021041,6802426}.
The two flux-qubit levels correspond to the clockwise and anti-clockwise circulating currents in the qubit loop (denoted `Q' in Fig.~\ref{fig:CCJJ}).

In the $\{\ket{\circlearrowright},\ket{\circlearrowleft}\}$ basis, the Hamiltonian of the CCJJ flux qubit is off-diagonal due to the tunneling between these states across the finite potential barrier. We, therefore, have two independent control parameters: $\epsilon$ that controls the asymmetry between two potential wells, and $\Delta$ that controls the potential barrier height between two wells. The Hamiltonian of this qubit is ~\cite{PhysRevB.82.024511,HJB+10}
$\hat{H}_{{\rm Q}}=-\left(\epsilon\sigma^z+\Delta\sigma^x\right)$
for $\epsilon=|I_{{\rm Q}}|\left(\Phi_{{\rm Q}}-\Phi_0\right)$
with~$I_{{\rm Q}}$ the super-current circulating in the rf-SQUID loop,
$\Phi_{{\rm Q}}$ the time-dependent qubit-flux bias applied across the qubit,
and $\sigma^{x,y,z}$ the three Pauli matrices.
It is possible to turn off the diagonal terms of the Hamiltonian by setting $\Phi_{{\rm Q}}=\Phi_0$, a regime often referred to as the \emph{degeneracy point}.
Tunneling energy~$\Delta$ is a function of $\Phi_{{\rm CCJJ}}$ that can be controlled via $I_{{\rm CCJJ}}$.

Fabricating such qubits having 
identical controllability but superior energy relaxation time (expected $T_1 > 10~\mu$s) is underway~\cite{Note2},
which could serve as an ideal platform for conducting such QPT experiments.
Coherence times of tens of microseconds are orders of magnitude higher than our operation time for simulating the QPT,
which guarantees that the effect of decoherence-induced noise is negligible for our case.
On the other hand, superconducting qubits operate at around $10$ mK~\cite{Niskanen04052007,0953-2048-23-4-045027,Note2},
which corresponds to a thermal excitation with frequency~$\sim 1.3$ GHz,
which is far detuned from the frequencies of the quasiparticles in the flux qubit ($\sim 6$ GHz.)~\cite{PhysRevX.4.021041}.
This detuning implies that the mean density of thermal photons in the system at the operating temperature is less than $1\%$, 
so the possibility of absorbing thermal excitations by the system is minimal,
thereby rendering our architecture resilient against possible environment-induced noises.

\begin{figure}[htb]
\centering
\includegraphics[width=0.7\linewidth]{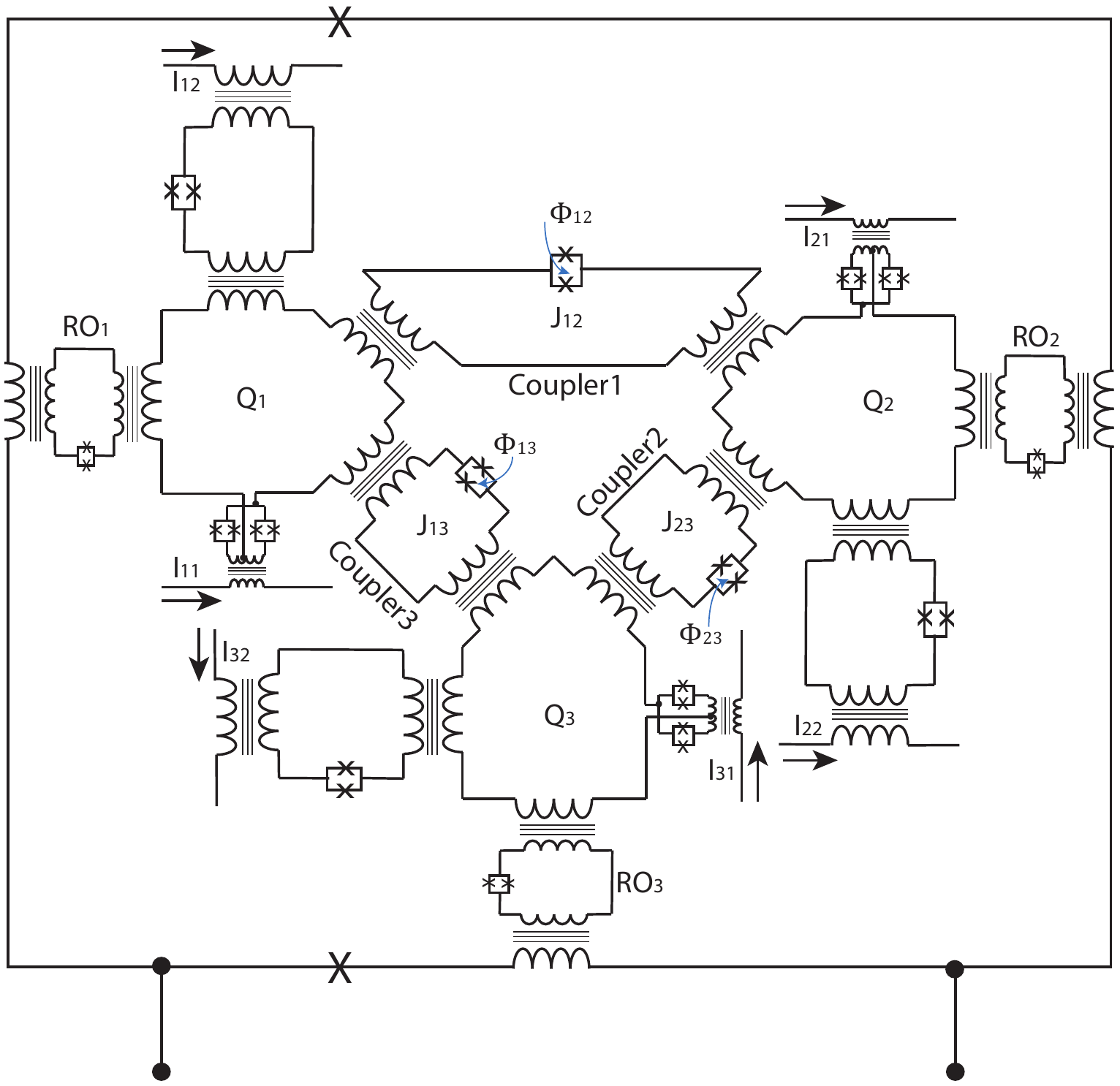}
\caption{A circuit diagram of three rf-SQUID flux qubits coupled with each other via inductive tunable couplers. The square around the system couples all three flux qubits and measure the total magnetic moment of the system which we use towards the macroscopic characterization of the QPT in the network.}
\label{fig:coupledCCJJ}	
\end{figure}

\subsection{Triangular architecture of coupled CCJJ qubits}
We now consider a triple-spin network 
with the three CCJJ flux qubits inductively coupled in a triangular arrangement shown in Fig.~\ref{fig:coupledCCJJ}.
In this architecture, each pair of qubits is coupled via an inductive coupler that can be tuned {\it in situ} via the external flux biases $\Phi_{ij}$~\cite{PhysRevB.82.024511}. Such an inductive coupling scheme induces a $\sigma_i^z\sigma_j^z$-type interaction between the $i^{{\rm th}}$ and the $j^{{\rm th}}$ qubits with an adjustable coupling strength~$J_{ij}$~\cite{PhysRevB.82.024511}.
The Hamiltonian for this network of three coupled flux qubits is
\begin{equation}
\label{eq:fullHamiltonian}
	\hat{H}_{{\rm s}}
		=\sum_{i=1}^3 \hat{H}_{{\rm Q}}^{(i)}
			+\sum_{i=1}^2 \sum_{j=i+1}^3J_{ij}\sigma_i^z\sigma_j^z
\end{equation}
with~$i,j$ denoting qubit indices and $\hat{H}_{{\rm Q}}^{(i)}$ the Hamiltonian for the $i^{{\rm th}}$ flux qubit. Using the control mechanisms described earlier, we can 
achieve $\epsilon_i=0$ by setting each external qubit flux bias $\Phi_{Q}^{(i)}$ 
at the degeneracy point of the $i^{{\rm th}}$ qubit.

\section{Simulating QPT from paramagnetism to frustrated magnetism}
\label{sec:simulation}

In this section we discuss how to control the quantum Hamiltonian (\ref{eq:fullHamiltonian}) in order to achieve the frustrated regime starting from the paramagnetic phase. We compute the fidelity susceptibility for the ground state of the system and explore its divergent behavior, which in fact denotes a QPT without any prior knowledge of any local order parameter. Finally we discuss how micro and macro signatures of the QPT can be extracted for our system.

\subsection{Control scheme}
In our approach to simulating frustrated antiferromagnetism, we initially prepare the ground state of the system Hamiltonian (\ref{eq:fullHamiltonian}) with $\epsilon_i=J_{ij}\approx0\;\forall i,j$
and then vary~$\Delta$ and~$J$ (assuming $\Delta_i=\Delta$ and $J_{ij}=J$ for all $i,j$) adiabatically 
via $\Phi^{(i)}_{{\rm CCJJ}}$ and $\Phi_{ij}$,
respectively. 
The control pulses for these parameters are designed
such that,
at $t=0$, $J/\Delta \ll 1$ ($\sim 10^{-6}$), and,
at $t=t_{{\rm final}}$, $J/\Delta \gg 1$ ($\sim 10^6$)~\cite{PhysRevX.4.021041,Ronnow25072014}.
The Hamiltonian ground state $\ket{g}$ at $t=0$ is
$\ket{g(t=0)}=\ket{+++}$~\cite{Note6},
which is separable.
During the system's adiabatic evolution towards the frustrated regime,
the three qubits become entangled,
and the system ground state at $t=t_{\rm final}$ is
(neglecting global phase)
$(\ket{\downarrow\downarrow\uparrow}+\ket{\downarrow\uparrow\downarrow}+\ket{\uparrow\downarrow\downarrow}-\ket{\downarrow\uparrow\uparrow}-\ket{\uparrow\downarrow\uparrow}-\ket{\uparrow\uparrow\downarrow})/\sqrt{6}$.

\subsection{Fidelity susceptibility}

Fidelity susceptibility offers a measure for QPT in absence of any prior knowledge of order parameters~\cite{PhysRevE.76.022101, PhysRevX.5.031007}. The fidelity susceptibility $\chi_{\rm F}$ quantifies the drastic change in the ground state of the quantum system during the phase transition, and is defined as
\begin{equation}
\chi_{\rm F}(\lambda)=-\frac{\partial^2 \ln{\left|{\braket{\Psi_{0}(\lambda)}{\Psi_{0}(\lambda+\epsilon)}}\right|}}{\partial \epsilon^2}\Bigg|_{\epsilon=0},
\end{equation} 
where $\Psi_{0}$ is (normalized) ground state wavefunction of the system and $\lambda$ is the control parameter, which is $J/\Delta$ for our case. It is possible to show that $\chi_{\rm F}(\lambda) \geq 0$ for all $\lambda$, and can also be expressed as~\cite{PhysRevX.5.031007}
\begin{equation}
\label{eq:chiFUsed}
\chi_{\rm F}(\lambda)=\left\langle{\frac{\partial \Psi_0}{\partial \lambda}}\Bigg|{\frac{\partial \Psi_0}{\partial \lambda}}\right\rangle-\left\langle{\Psi_0}\Bigg|{\frac{\partial \Psi_0}{\partial \lambda}}\right\rangle \left\langle{\frac{\partial \Psi_0}{\partial \lambda}}\Bigg|{\Psi_0}\right\rangle.
\end{equation}
We use Eq.~(\ref{eq:chiFUsed}) to compute the $\chi_{\rm F}$ for our system, and seek its divergent behavior that denotes the QPT.

\begin{figure}[htb]
\centering
\includegraphics[width=0.9\linewidth]{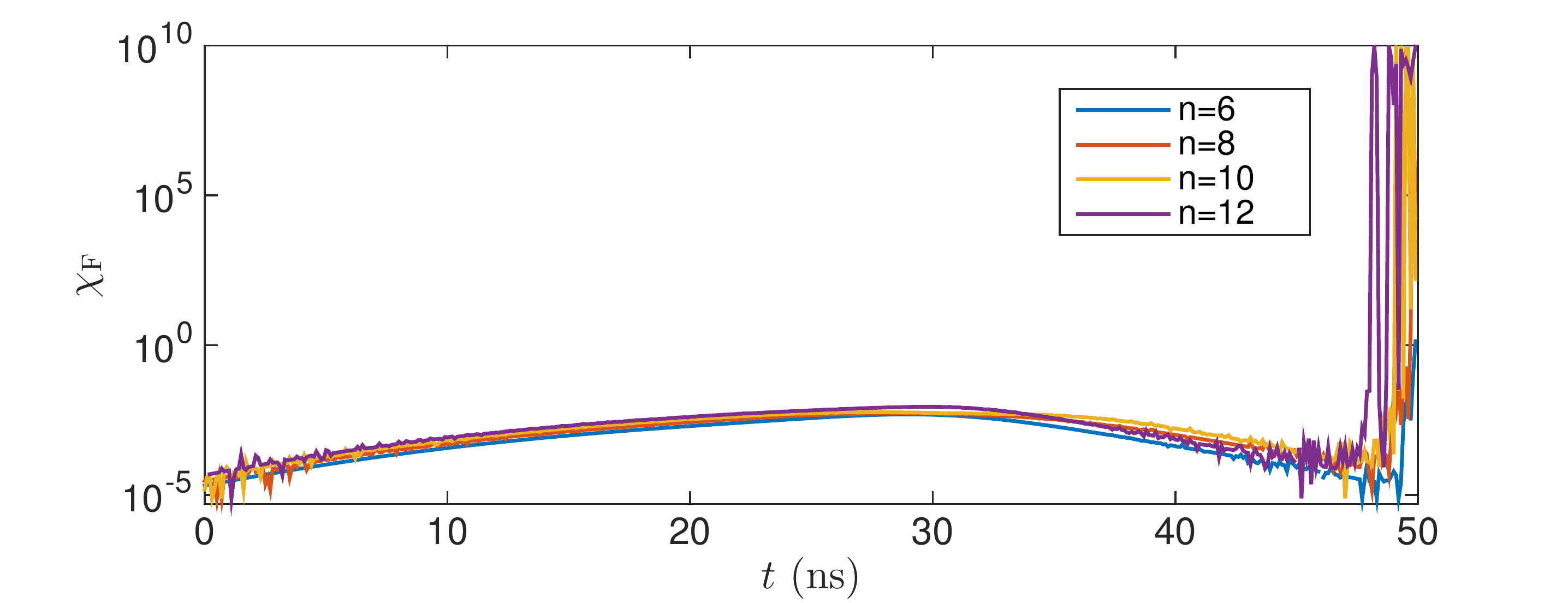}
\caption{Plots showing the fidelity susceptibility $\chi_{\rm F}$ as a function of time for $n$ nearest- and next-nearest-neighbor-coupled flux qubits. The divergence in $\chi_{\rm F}$ denotes a QPT.}
\label{fig:chiF}
\end{figure}

Fig.~\ref{fig:chiF} shows the fidelity susceptibility $\chi_{F}$ as a function of time for various lattice sizes. During the time $t=0$ to $t=50$ ns, we simultaneously vary $J$ from $0$ to $5$ GHz and $\Delta$ from $5$ to $0$ GHz adiabatically. We observe that while $\chi_{\rm F} \ll 1$ for most of the time during the change of our control parameter, it however shows divergence when $J/\Delta \gg 1$. The plots in Fig.~\ref{fig:chiF} not only indicates the existence of a QPT for our system, but also shows that extracting the signatures of QPT should be possible for a lattice with $n \approx 12$ for which the divergence of $\chi_{\rm F}$ is prominent. This requirement is especially useful as the classical simulation of any quantum spin system becomes exponentially expensive with higher values of $n$. For the macroscopic characterization of QPT, we therefore restrict ourselves to a lattice with $n=12$, which turns out to be sufficient for the architecture considered in this work.

\subsection{Microscopic signature}
QPTs are usually characterized microscopically by the divergence of correlation length near the critical point~\cite{sach2011}. Experimentally, however, it is not always efficient to determine this divergence, and, therefore, alternative system-specific schemes are often used, such as measurement of some local entanglement-witness operator for spin-systems~\cite{kim2010quantum}.

The QPT for our architecture can be characterized by selecting a block of three adjacent flux qubits (which is in fact a triangular architecture as shown in Fig.~\ref{fig:frustrationScheme}) and measuring a three-qubit entanglement-witness operator on that block.
Such a measurement is performed locally on a few-spin subsystem of the entire lattice. 
Moreover, the measurement of entanglement-witness operator requires readout operations on each individual qubit,
hence is a microscopic characterization.

The symmetric $W$-state witness operator~\cite{2009PhR4741G,kim2010quantum}
\begin{equation}
	W_{\rm S}
		:=4+\sqrt{5}-\frac{1}{2}\left(\sum_{i=1}^3\sigma_i^x\right)^2
			-\frac{1}{2}\left(\sum_{i=1}^3\sigma_i^{y}\right)^2,
\end{equation}
satisfies~$\langle{W_{\rm S}}\rangle>0$ for separable states and $\langle{W_{\rm S}}\rangle<0$ for entangled states.
For our case, we expect~$\langle{W_{\rm S}}\rangle=\sqrt{5}-2$ at $t=0$ and $\langle{W_{\rm S}}\rangle=\sqrt{5}-3$ at $t=t_{\rm final}$,
which prominently signifies the entanglement change.

\subsection{Macroscopic signature}
For macroscopic characterization, we can place a dc-SQUID loop around the entire system, such that it couples all the flux qubits (Fig.~\ref{fig:coupledCCJJ} shows such a loop for the three-qubit case)
and measure the total magnetic moment $\hat{\mu}$ along the $z$-axis~\cite{Note4}. 
The net magnetic moment of the entire system along $z$-axis ${\hat{\mu}^z}$ 
can be expressed as a sum of the $z$-components of the individual spin magnetic moments 
$\sum_{i=1}^{n}\sigma_i^z$ of the entire system.
Classically, the total magnetic moment of a system per unit volume is the magnetization of the system, which is a macroscopic quantity.
We refer to this readout procedure as a \emph{macroscopic} characterization of frustration,
as it does not measure any local properties of the system addressing each flux qubit individually (elaborated in~\ref{sec:KLD}). 

The spectrum~$\{\mu^z\}$ of the total magnetic moment operator~${\hat{\mu}^z}$ assumes only integer values.
Repeated measurements yield a probability distribution for~$\{\mu^z\}$
centered at~$\mu^z=0$ for both the frustrated as well as paramagnetic phases.
The tail of the distribution
differs between frustrated and paramagnetic phases.
For the paramagnetic phase, the state of the entire system is
$\ket{+}^{\otimes{n}}$, which indicates that all states
are equally probable, if measured in the $\{\ket{\uparrow},\ket{\downarrow}\}^{\otimes n}$ basis.
Therefore, the probability that $l$ spins are in $\ket{\uparrow}$ state 
and $l+k$ spins (with $n=2l+k$) are in $\ket{\downarrow}$ state (for which $|\mu^z|=k$), is 
$\frac{1}{2^n} {n \choose (n-k)/2}{n \choose (n+k)/2}$, which means that the 
probability distribution of~$\mu^z$ is a binomial function for the paramagnetic case.
For the frustrated phase, the probability distribution depends on the geometry of the lattice, 
which is why it is called geometric frustration. Whereas, an exact closed-form analytic solution 
of the distribution for this case is hard to derive, the frustrated antiferromagnetism implies that 
the peak of the distribution is still centered at~$\mu^z=0$,
and the tail is an exponential decay (as opposed to binomial), which can be verified by numerical simulations.

In order to characterize the change in the tail
of the probability distribution, we introduce a measure
\begin{equation}
	\mathcal{D}[\mathcal{P}(t)]:=D_{{\rm KL}}[P_{{\rm exp}},\mathcal{P}(t)]-D_{{\rm KL}}[P_{{\rm bin}},\mathcal{P}(t)],
\label{eq:D}
\end{equation} 
on the probability distribution $\mathcal{P}(t)$ at time~$t$
with~$P_{{\rm exp}}$ and~$P_{{\rm bin}}$ best-fit exponential and binomial distributions corresponding to the final and initial states.
The Kullback-Leibler divergence~$D_{{\rm KL}}:=\sum_kP_k\log({P_k/Q_k})$
quantifies distinguishability between two probability distributions $P$ and $Q$.
If the distribution~$\mathcal{P}(t)$ changes from binomial to exponential,
then~$\mathcal{D}$ goes from positive to negative,
which indicates a phase transition from paramagnetic to frustrated magnetic phase has occurred.

\section{Results}
\label{sec:results}

\begin{figure}[htb]
\centering
\includegraphics[width=0.7\linewidth]{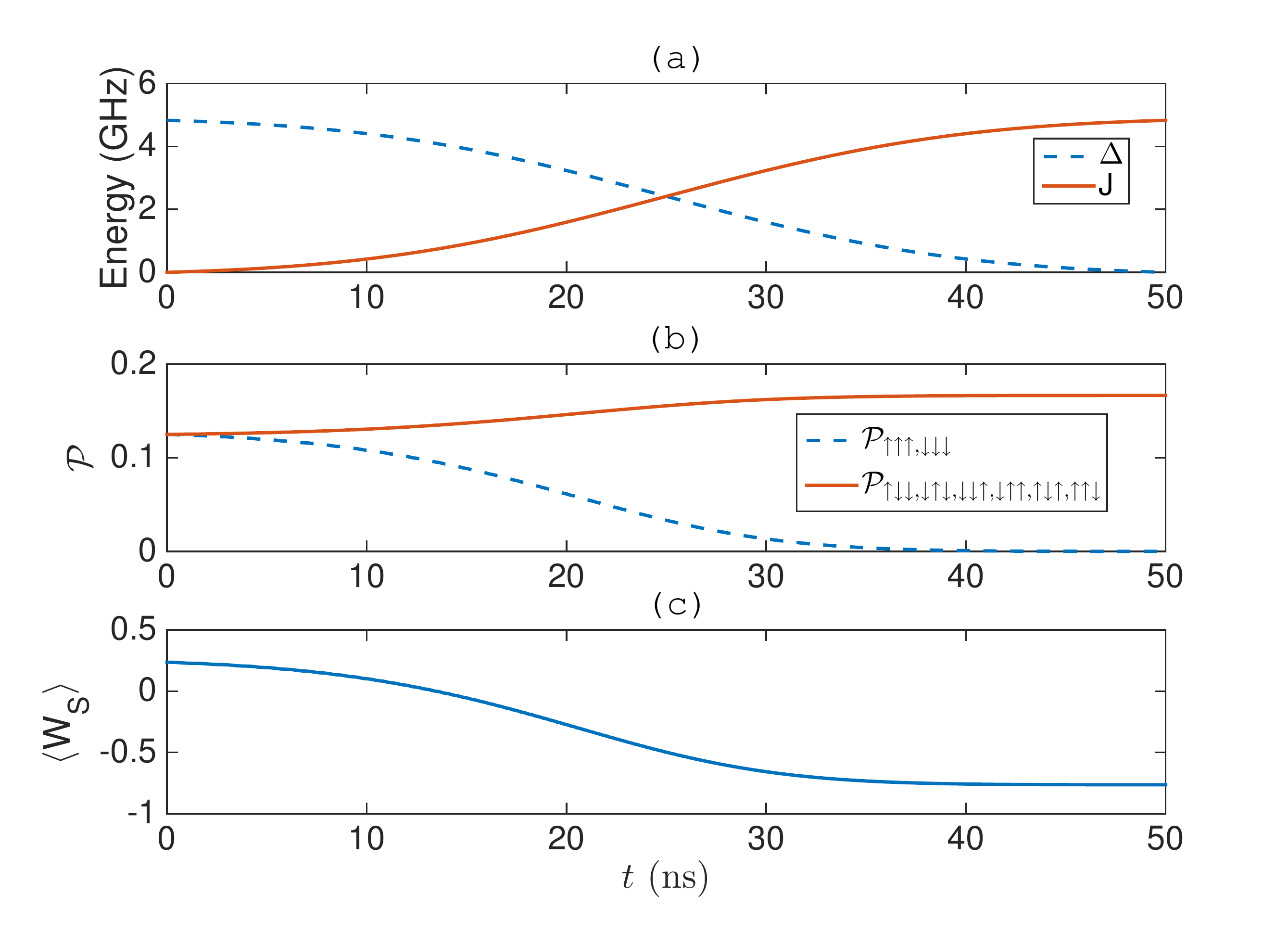}
\caption{Plots showing the microscopic signatures of QPT. (a) The control pulse where $\Delta$ gets turned off and~$J$ turned on adiabatically with time. (b) The probability~$\mathcal{P}$ for each state (denoted by subscripts in the legend) with time under the control pulse. (c) Expectation of the symmetric $W$-state witness operator as a function of time under the control pulse. The change of sign indicates the QPT.}
\label{fig:frustrationMicro}
\end{figure}

Figure~\ref{fig:frustrationMicro} shows the results for the microscopic characterization of the QPT for our three-qubit device. The adiabatic pulses for $\Delta$ 
and~$J$ are shown in Fig.~\ref{fig:frustrationMicro}(a)
for couplings slowly turned on and local controls off.
Figure~\ref{fig:frustrationMicro}(b) shows the probability distribution for all eight states.
Whereas at $t=0$ the probabilities are all equal,
they gradually decay for~$\ket{\uparrow\uparrow\uparrow}$ and~$\ket{\downarrow\downarrow\downarrow}$ states with a uniform probability distribution for the remaining states, which is expected for a paramagnetic to frustrated magnetic phase transition. Figure~\ref{fig:frustrationMicro}(c) shows the change of symmetric $W$-state entanglement-witness operator with time. At $t=t_{{\rm final}}$ the state of the system is a superposition of two $W$ states, and therefore $\langle{W_{S}}\rangle$ goes from positive to negative indicating the generation of entanglement in the system, which we consider as our microscopic signature.

\begin{figure}
\centering
\subfloat[]{
\centering
\includegraphics[width=0.7\linewidth]{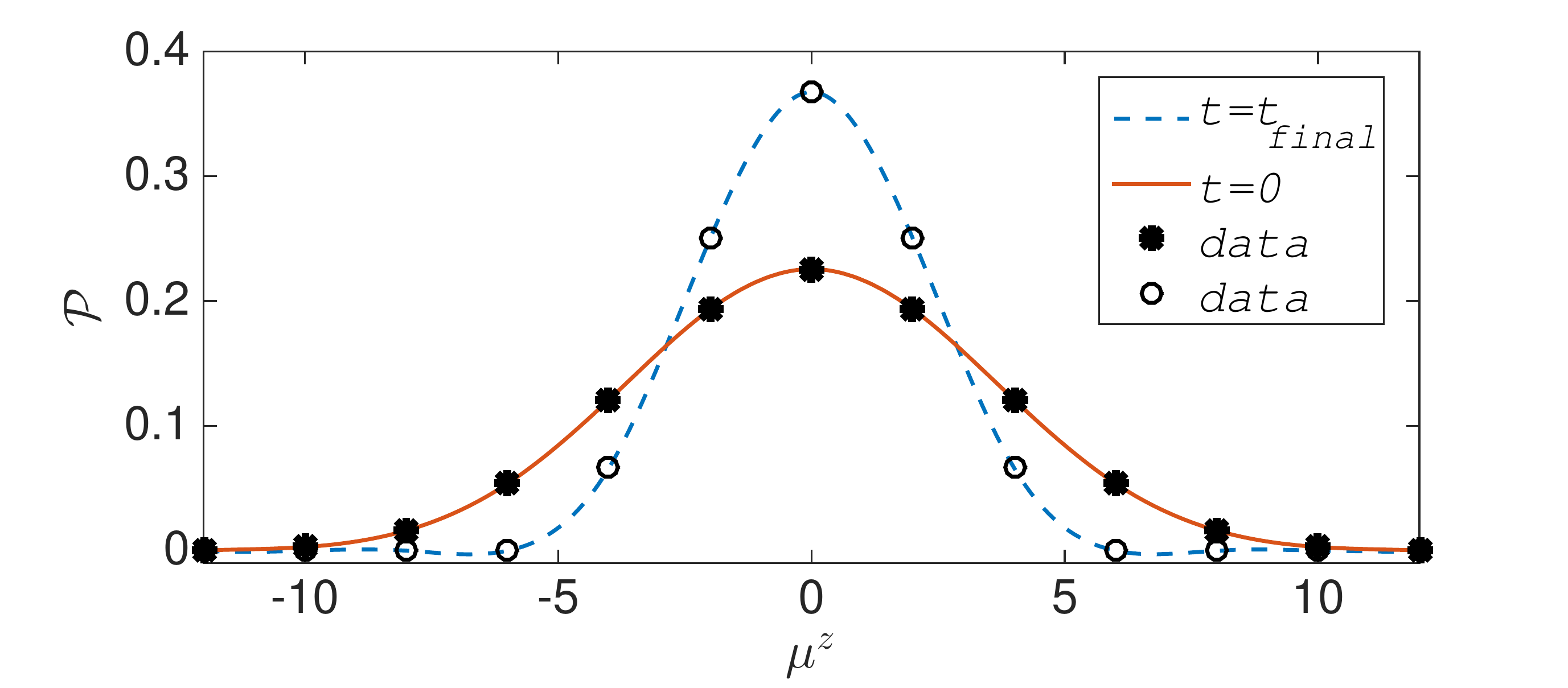}
\label{fig:frustrationMacroProb}
} \\
\subfloat[]{
\centering
\includegraphics[width=0.7\linewidth]{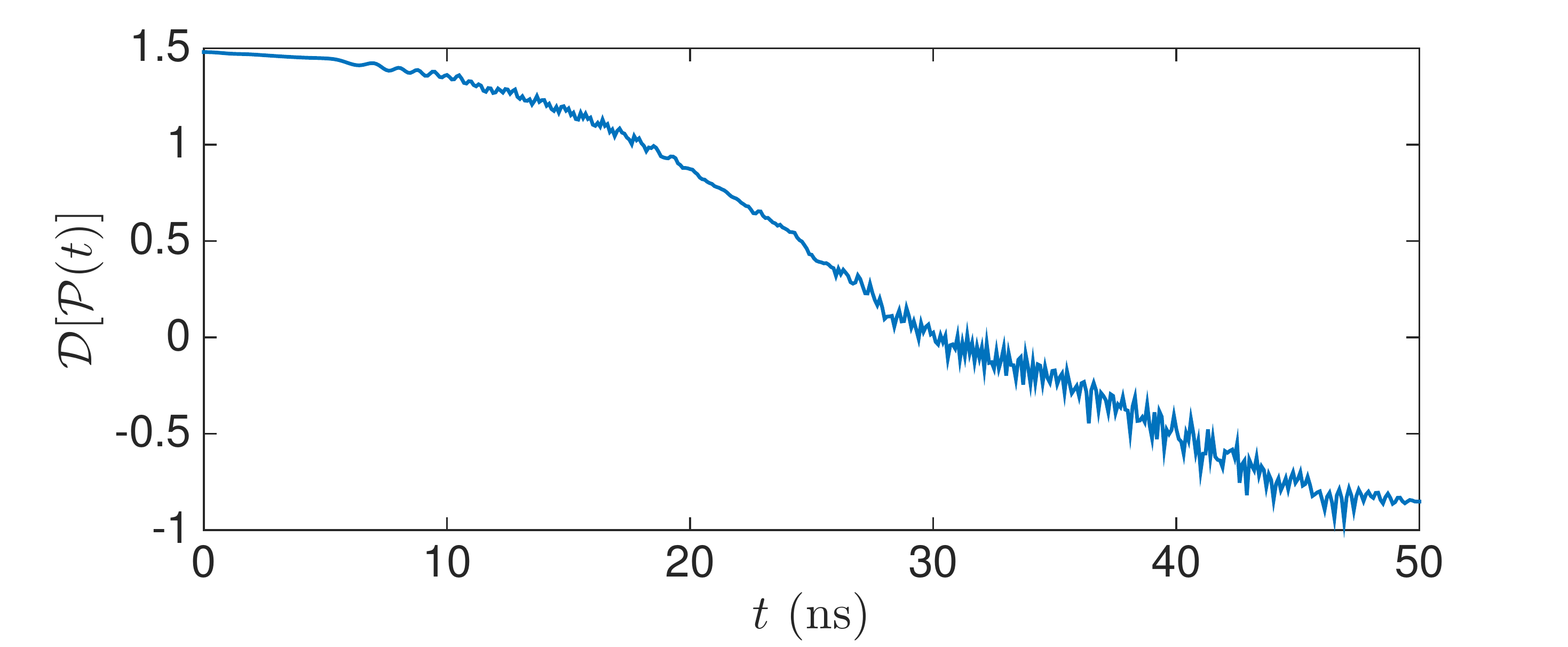}
\label{fig:frustrationMacroD}
}
\caption{%
	Macroscopic signatures of the QPT. (a) The probability distributions $\mathcal{P}(\mu^z)$ of the total magnetic moment~$\mu^z$ for the network of $12$ flux qubits at initial and final times. The black dots denote the actual data computed via numerical simulation and solid lines denote interpolations through the data. The tail changes from binomial to exponential denoting the QPT. (b) Our distance measure based on the Kullback-Leibler divergence as defined in Eq.~(\ref{eq:D}) as a function of time.
	The change in sign indicates the QPT.%
	}
\label{fig:frustrationMacro}
\end{figure}

Figure~\ref{fig:frustrationMacro} shows the macroscopic signatures of the QPT for an $n$-qubit network of nearest-neighbor- and next-nearest-neighbor-coupled superconducting flux qubits
shown in Fig.~\ref{fig:frustrationChain}. We plot the initial and final probability distributions for the total magnetic moment of the system in Fig.~\ref{fig:frustrationMacroProb} for $n=12$. Although for both cases the distribution is centered around the origin, a significant difference is observed in the tails. The tail of the probability distribution changes from a binomial to an exponential decay. $\mathcal{D}[\mathcal{P}(t)]$ quantifies this change and \ref{fig:frustrationMacroD} shows how~$\mathcal{D}$ changes gradually from positive to a negative value with time signifying a QPT with respect to a macroscopic probe, which is the SQUID loop around the network for our case.

\section{Conclusions}
\label{sec:conclusions}
In this work, we have put forward a realistic scheme to simulate a QPT, where a network of nearest-neighbor- and next-nearest-neighbor-coupled superconducting flux qubits is subjected to a transition from the paramagnetic phase to the frustrated magnetic phase. 
Characterizing a QPT with a micro as well as a macro probe is a conceptual requirement,
rather than an experimental preference.

Whereas the microscopic characterization of the phases can be performed measuring the appropriate entanglement witness, we have introduced a measure in this work based on the Kullback-Leibler divergence that can characterize the QPT macroscopically. 
Our proposal for such simultaneous characterization of QPTs at macro and micro scales
enables a rigorous authentication of various quantum phases, 
as well as offers a standardized procedure to self-consistently verify 
the signatures of quantum phase transitions in upcoming experiments.

\section*{Acknowledgments}
This research was funded by NSERC and AITF.
JG acknowledges support from the University of Calgary's Eyes High Fellowship Program,
and appreciates support from China's Thousand Talent Plan.
We thank A.~Blais, M.~Geller,
D.~Home, M.~Saffman and P.~Roushan for valuable discussions.

\appendix

\section{Kullback--Leibler Divergence: A Macroscopic Measure for QPT}
\label{sec:KLD}
A phase transition, classical or quantum, is always defined for a macroscopic system. Here we are interested in the question if the phase transition can be characterized macroscopically or not; in other words, if the probe for the QPT is macroscopic or not. The motivation for this question is described in the previous section, and here we elaborate on why the SQUID loop considered for characterizing the transition from paramagnetic to frustrated magnetic phase is in fact macroscopic.

The SQUID loop around the entire network of nearest- and next-nearest-neighbor-coupled flux qubits is measuring the total magnetic moment $\hat\mu^{z}$ of the system, which in terms of the individual spins can be expressed as, $ {\hat\mu^{z}}=\sum_{i=1}^{n}\sigma_{i}^{z}$. However the SQUID loop does not resort to measuring each individual spins and then summing it up; in fact it does not even have any access to each individual spins in the network. Therefore, this probe functions as a macroscopic probe.

Since the spins in the network have many possible arrangements, the SQUID loop can find different values for the total magnetic moment of the system corresponding to various spin-arrangements, and by repetitive measurements one can construct the probability distribution for these different arrangements. As the control parameters are varied from the paramagnetic regime to the frustrated magnetic regime, the probability distribution also changes, and the question of characterizing the phase transition boils down to the question of how to distinguish between two probability distributions quantitatively.

The probability distributions corresponding to paramagnetic and frustrated magnetic phases are shown in Fig. 5. The peaks for both of these probability distributions are centered about the origin ($|\mu^{z}|=0$), whereas their tails differ significantly. We thus point out that the tail of the probability distribution carries the information about the phase. In order to quantify the change in the tail of the distribution we propose to compute the Kullback-Leibler divergence $D_{{\rm KL}}$, a non-symmetric measure of the distance between two probabilistic distributions. It is important to mention in this context that the measures like Kullback-Leibler divergence or Jensen-Shannon divergence has already been used to characterize the classical phase transition (not as a macroscopic measure) from Bose gas to Bose-Einstein condensate, where the transition is also characterized by the change in the tail of the velocity-distribution~\cite{CH.CHATZISAVVAS2006}.

The measure we introduced here to characterize the QPT (given by Eq. (3)) takes positive value for paramagnetic and negative value for frustrated magnetic phase, and thereby quantifies not only the change in phase, but also indicates the phase itself for the entire system. It is important to note that the Kullback-Leibler divergence is a measure of classical relative entropy, whereas characterizing a QPT for our case. This is not surprising, since the change in the probability distribution for different phases are still mediated by quantum fluctuations, and a classical measure is sufficient to quantify this change. One can possibly introduce some macroscopic measure of entanglement contained in the entire network, and characterize the QPT based on that measure. However, we do not attempt to define such a measure as it does not change the conclusions of our work.

\section*{References}
\bibliography{frustration}

\end{document}